\documentclass[aps,prl,twocolumn,showpacs,superscriptaddress]{revtex4}  
\usepackage{amssymb,bm,graphicx,mathptmx,color,dsfont}
\RequirePackage[utf8]{inputenc} 

\begin{document}

\title%
{Generation of continuous-variable cluster states of cylindrically polarized modes}

\author{Ioannes~Rigas}
\thanks{These authors contributed equally to this work.} 
\affiliation{Max Planck Institute for the Science of Light, Staudtstr.~2, D-91058 Erlangen, Germany} 
\affiliation{Institute of Optics, Information and Photonics, University Erlangen-Nuremberg, Staudtstr.~7/B2, D-91058 Erlangen, Germany}

\author{Christian~Gabriel}
\thanks{These authors contributed equally to this work.} 
\affiliation{Max Planck Institute for the Science of Light, Staudtstr.~2, D-91058 Erlangen, Germany} 
\affiliation{Institute of Optics, Information and Photonics, University Erlangen-Nuremberg, Staudtstr.~7/B2, D-91058 Erlangen, Germany}

\author{Stefan~Berg-Johansen}
\affiliation{Max Planck Institute for the Science of Light, Staudtstr.~2, D-91058 Erlangen, Germany} 
\affiliation{Institute of Optics, Information and Photonics, University Erlangen-Nuremberg, Staudtstr.~7/B2, D-91058 Erlangen, Germany}

\author{Andrea~Aiello}
\affiliation{Max Planck Institute for the Science of Light, Staudtstr.~2, D-91058 Erlangen, Germany} 
\affiliation{Institute of Optics, Information and Photonics, University Erlangen-Nuremberg, Staudtstr.~7/B2, D-91058 Erlangen, Germany}

\author{Peter~van~Loock}
\affiliation{Max Planck Institute for the Science of Light, Staudtstr.~2, D-91058 Erlangen, Germany} 
\affiliation{Institute of Optics, Information and Photonics, University Erlangen-Nuremberg, Staudtstr.~7/B2, D-91058 Erlangen, Germany}
\affiliation{Institute of Physics, University of Mainz, Staudingerweg 7, 55128 Mainz, Germany}

\author{Ulrik~L.~Andersen}
\affiliation{Max Planck Institute for the Science of Light, Staudtstr.~2, D-91058 Erlangen, Germany} 
\affiliation{Institute of Optics, Information and Photonics, University Erlangen-Nuremberg, Staudtstr.~7/B2, D-91058 Erlangen, Germany}
\affiliation{Department of Physics, Technical University of Denmark, 2800 Kongens Lyngby, Denmark}

\author{Christoph~Marquardt}
\affiliation{Max Planck Institute for the Science of Light, Staudtstr.~2, D-91058 Erlangen, Germany} 
\affiliation{Institute of Optics, Information and Photonics, University Erlangen-Nuremberg, Staudtstr.~7/B2, D-91058 Erlangen, Germany}
\email{Christoph.Marquardt@mpl.mpg.de}

\author{Gerd~Leuchs}
\affiliation{Max Planck Institute for the Science of Light, Staudtstr.~2, D-91058 Erlangen, Germany} 
\affiliation{Institute of Optics, Information and Photonics, University Erlangen-Nuremberg, Staudtstr.~7/B2, D-91058 Erlangen, Germany}

\begin{abstract}
Cluster states are an essential component in 
one-way quantum computation protocols. 
We present two schemes to generate addressable continuous-variable cluster
states from quadrature squeezed cylindrically polarized modes. 
By including polarization in addition to the transverse spatial degree of
freedom, elementary cluster states can be created in which four cluster nodes
co-propagate within one paraxial vector beam. This approach is fundamentally
compatible with existing time-multiplexed schemes that have been used to create
the largest cluster states to date.
We implement a proof-of-principle experiment of one of the
proposed schemes and verify its feasibility by measuring the quantum
correlations between the different nodes of the cluster state.
\end{abstract}
\maketitle

The efficient processing of computational tasks is extremely desirable in
today's fast-evolving information society. Quantum computers have the potential
to clearly outperform classical systems at certain computational operations  
\cite{bennett-quantum-2000}. 
While recent years have seen a steady development in the technologies
necessary for full-scale quantum computing, work continues also on foundational
concepts and new implementation candidates.
It has been shown that so-called cluster states represent a universal resource
for measurement-based, one-way quantum computation, which can be performed
with discrete-variable (DV) systems
\cite{%
raussendorf-one-way-2001,
raussendorf-measurement-based-2003,
nielsen-optical-2004,
walther-experimental-2005,
van-den-nest-universal-2006} 
as well as with continuous-variable (CV) quantum states
\cite{%
menicucci-universal-2006,
zhang-continuous-variable-2006,
su-experimental-2007,
menicucci-ultracompact-2007,
van-loock-building-2007,
menicucci-one-way-2008,
pysher-parallel-2011,
menicucci-graphical-2011,
wang-weaving-2014,
menicucci-fault-tolerant-2014,
roslund-wavelength-multiplexed-2013,
alexander-flexible-2016}. 
In these schemes specially prepared entangled states, the cluster states, form
the backbone of the protocol. The central idea behind cluster-state quantum
computation is to enact quantum logic gates on parts of an initially prepared
cluster state by teleporting them through the cluster.
These gates are then implemented solely by means of local measurements, while additional
Hadamard (DV) or Fourier (CV) gates are built-in along the teleportations to
switch between different bases and obtain universality. 
The first measurement results determine which bases to choose
for the remaining measurements and which Pauli (DV) or displacement (CV)
corrections to apply.
Thus, in this model, the complication of implementing deterministic quantum
gates is transferred to the task of generating highly non-locally correlated
cluster states, typically, either in a probabilistic (DV) or deterministic (CV)
fashion~\cite{andersen-hybrid-2015}.
 
There has been significant progress in generating large scale CV cluster
states in recent years. The first realization of CV cluster states relied on
the generation of individually squeezed states from different optical
parametric oscillators supporting the same Gaussian spatial modes. The modes
interfere in a linear network of beam splitters to create a cluster state of
four modes~\cite{su-experimental-2007,yukawa-experimental-2008}. 
This idea was then extended to higher order spatial modes combined with
a programmable spatial mode detector rendering the system much more
compact~\cite{armstrong-programmable-2012}. These methods did
not allow for large scale cluster states to be formed. However, excellent
scalability can be attained by making use of the inherent quantum correlations
between frequency modes of an optical parametric
oscillator~\cite{menicucci-ultracompact-2007,menicucci-one-way-2008,pysher-parallel-2011,pinel-generation-2012} or by using the temporal-mode quantum
correlations~\cite{menicucci-arbitrarily-2010,menicucci-temporal-mode-2011}.
Using these two approaches, cluster states of 16 modes~\cite{roslund-wavelength-multiplexed-2013}, 60
modes~\cite{chen-experimental-2014} and 10$^6$ modes~\cite{yokoyama-ultra-large-scale-2013,yoshikawa-invited-2016} have
been generated, respectively.         
  
Here, we report on new schemes to generate a four-state cluster 
with the help of quadrature squeezed cylindrically polarized modes.
These modes have a nonseparable structure leading to useful properties 
both in a quantum~\cite{gabriel-entangling-2011} and classical optics
setting~\cite{aiello-quantumlike-2015}.
The resulting cluster states rely on the four orthogonal bases
of the cylindrically polarized modes which can be easily distinguished and
addressed.
This approach to cluster generation is related to the approach of
Ref.~\cite{armstrong-programmable-2012} where the different eigenmodes of the cluster
correspond to the higher-order Hermite-Gaussian spatial modes. However, in our
proposal, we add the polarization as another degree of freedom, thereby
extending the mode space by a factor of two and promoting the flexibility in
addressing the various modes. The approach can be extraordinarily compact as
all involved modes can be generated in a single optical parametric oscillator
using the process of parametric
downconversion~\cite{dos-santos-continuous-variable-2009,lassen-continuous-2009} or
by using the optical Kerr effect of an optical fiber~\cite{euser-birefringence-2011,
gabriel-entangling-2011}. 
The schemes we propose here are designed in such a way that they offer,
at the same time, a compact and robust cluster-state preparation
as well as an easy and direct addressability of the nodes of
the cluster. At the end of this paper, we report on a proof-of-principle 
demonstration of one of the proposed schemes, for which we measure
the quantum correlations present among all the four nodes of the cluster
state.
\begin{figure*}
\begin{center}
\includegraphics[width=16cm]{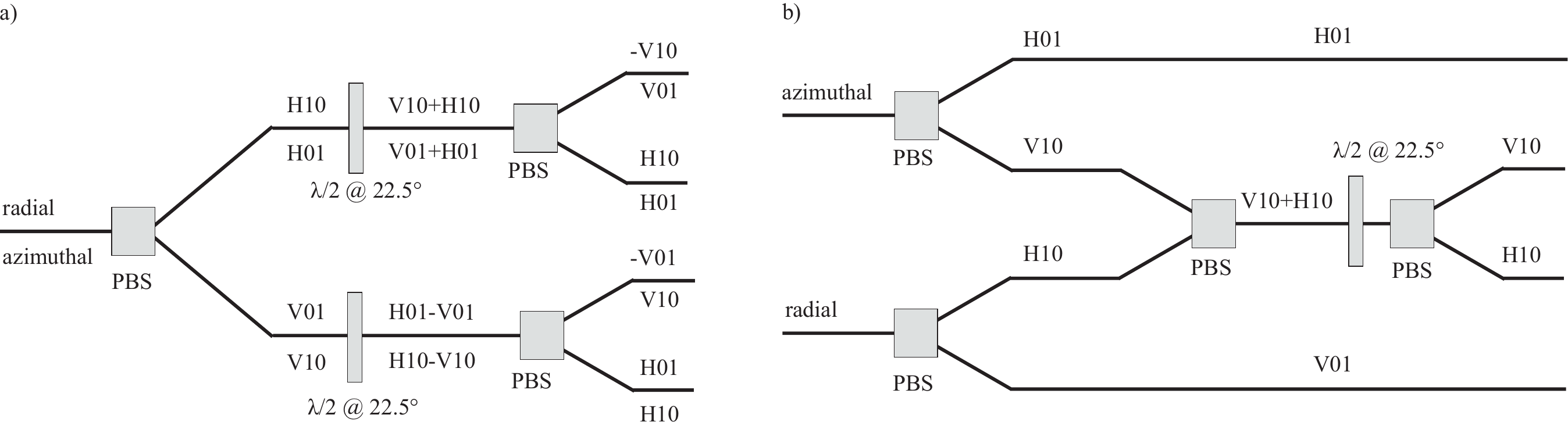}
\caption{\label{fig:Schemes} a) Scheme 1 for cluster-state
  generation. A radially polarized mode enters a 
  linear-optics network with the four output modes then forming a cluster
  state. The radially polarized mode could also be substituted by an
  azimuthally polarized one to generate a similar cluster state. 
  b) Scheme 2 for cluster-state generation. One radially and one azimuthally
  polarized mode form the input state. The four output modes again
  constitute a cluster state. [polarizing beam splitter (PBS), half-wave
  plate
  ($\lambda/2)$].}
\end{center}
\end{figure*}

For continuous variables, in the limit of infinite squeezing, cluster
states fulfill the eigenequation \cite{van-loock-building-2007}
\begin{equation}
  \label{eq:GenericCvClusterA}
\Big(\hat p_j - \sum_{k \in N_j} A_{jk} \hat q_k \Big)|\psi\rangle \to
0\quad \forall j. 
\end{equation}
This is in analogy to the discrete case, where cluster states are
eigenstates
of the Pauli stabilizers $\hat \sigma_x^{(j)} \prod_{k \in N_j} \hat
\sigma_z^{(k)}  $  \cite{briegel-persistent-2001}. In the CV case,
$\hat p_j$ and $\hat q_j$ correspond to
the ``position'' and ``momentum'' operators of the optical mode
$j$. $N_j$ are the adjacent modes of the mode $j$.
The real matrix $\mathbf{A}$ contains the full information about the
cluster or, equivalently, its graph. It therefore determines what
the displacement corrections after measuring some of the modes
must be in order to achieve, in principle, unit-fidelity teleportation.
This is reflected by the fact that the
covariances of the operators $(\hat p_j - \sum_k A_{jk}\hat q_k)$,
which correspond to the 
excess noises acquired during the individual teleportation
steps, all vanish. Note that in a two-mode scenario,
the state that satisfies Eq.~(\ref{eq:GenericCvClusterA})
is the famous EPR state up to a local Fourier rotation.

In any realistic scenario, however, one is limited to finitely squeezed input
states. Thus, quantum teleportation using
realistic cluster states always has non-unit fidelities, so that these physical
clusters are non-ideal states for universal quantum
computation. Nevertheless, 
one can define
\textit{approximate}
cluster states or \textit{Gaussian graph states} \cite{menicucci-graphical-2011}
which clearly specify the correlations between the individual modes in
a cluster, as well as the
infidelities of the quantum operations that have to be accepted for
a finite degree of squeezing.
The full information on any (zero-mean) Gaussian state is contained
in its covariance matrix.
CV cluster states can then be completely described by
the adjacency matrices of their graphs
\cite{menicucci-graphical-2011},
defined as,
\begin{equation}
\mathbf{Z} = i\mathbf{U} + \mathbf{V}.
\end{equation}
Here, $\mathbf{U}$ and $\mathbf{V}$ are given by
\begin{equation}
  \label{eq:DefUandV}
  \mathbf{U} = \frac{1}{2}\langle
\mathbf{\hat q}\mathbf{\hat q}^T\rangle^{-1},\quad
\mathbf{V} = \langle
\mathbf{\hat q}\mathbf{\hat q}^T\rangle^{-1} \times
\langle  \{   \mathbf{\hat
  q},\mathbf{\hat  p}^T\} \rangle,
\end{equation}
where $\langle \mathbf{\hat q}\mathbf{\hat q}^T\rangle$ denotes the
position covariance matrix and $\langle  \{   \mathbf{\hat
  q},\mathbf{\hat  p}^T\} \rangle$ the matrix describing the
cross-correlations between position and momentum, defined as  $\langle
\{   \mathbf{\hat  q},\mathbf{\hat  p}^T\} \rangle_{jk} =
\frac{1}{2}\langle \left(\hat q_j \hat p_k + \hat p_k\hat q_j \right)
\rangle$. The role of $\mathbf{U}$ is to incorporate the finite
squeezing of each mode, while $\mathbf{V}$ represents the coupling
between the modes. For the complex adjacency matrix $\mathbf{Z}$,
replacing the real matrix $\mathbf{A}$,
Eq.~(\ref{eq:GenericCvClusterA}) becomes an exact zero-eigenequation
even for finite squeezing \cite{menicucci-graphical-2011}. It is
therefore a convenient tool to quantify and describe finitely
squeezed cluster states.

For a realistic cluster with finite squeezing, the complex adjacency matrix
now determines both the necessary displacements during teleportation
and the 
excess noise that inevitably will be added to the state after teleportation:
the covariances of the operators $(\hat p_j - \sum_k
V_{jk}\hat q_k)$ are given by the matrix elements $U_{nm}/2$, so that
the total 
matrix $\mathbf{Z}$ contains the full information on the intermode
correlations of a cluster state.

Our cluster-state generation schemes are based on cylindrically
polarized beams.  These can be described by a superposition of
two out of four distinct modes. If one chooses, for example, the
Hermite-Gaussian 
basis to define cylindrically polarized modes, this would be the two
first-order Hermite Gaussian $\mathrm{TEM}_{10}$ (10) and
$\mathrm{TEM}_{01}$ (01) with horizontal (H) and vertical (V)
polarization, namely $H10$, $V10$, $H01$ and $V01$
\cite{holleczek-classical-2011}. In Ref.~\cite{gabriel-entangling-2011} we have shown that by
quadrature squeezing a cylindrically polarized beam, one generates
squeezing in and  entanglement between the basis modes. Our cluster states rely
precisely on these properties. The two most common cylindrically
polarized modes are the radially and azimuthally polarized modes, described by
the annihilation operators $\hat a_R$ and $\hat a_A$, respectively. Their
annihilation operators are defined as follows,
\begin{equation}
  \label{eq:CylPol}
  \hat a_R = \frac{1}{\sqrt{2}} ( \hat a_{H10} + \hat a_{V01} ) ,
  \quad
 \hat a_A = \frac{1}{\sqrt{2}} ( \hat a_{V10} - \hat a_{H01} ).
\end{equation}
The entanglement-generating operation is modeled by
the squeeze operator
\begin{equation}
  \label{eq:Squeezer}
  \hat S(r,\theta) = \exp[r (e^{-i\theta} \hat a^2 - e^{i\theta}\hat a^{\dag 2} )/2],
\end{equation}
applied before the beam enters the passive circuit which generates the final
cluster state.

In the present work, we propose two schemes to generate two different
types 
of cluster states.  A key element in both schemes are, on the one
hand, polarizing beam
splitters (PBSs), which act as mode separators and combiners and, on
the other hand, half-wave plates orientated at $22.5^\circ$. These
rotate the
input state by $\hat a_{H/V} \mapsto (\hat a_H \pm \hat
a_V)/\sqrt{2}$, thus performing a mode mixing between the different
input modes.

The first scheme is illustrated in
Fig.~\ref{fig:Schemes}a). A quadrature squeezed azimuthally (or
radially)  polarized mode is sent through an array of PBSs  and
half-wave plates. These are configured in such a
way that one gains access to all four basis modes which are spatially
separated at the
output of the circuit. In Scheme 2, displayed in
Fig.~\ref{fig:Schemes}b),   quadrature squeezed azimuthally and
radially polarized modes are split into their basis modes. In the next
step one pair of orthogonally polarized modes is combined on a PBS and
then mixed with a half-wave plate orientated at $22.5^\circ$. After
the half-wave plate, the modes are separated again at a PBS,
therefore allowing one to access all four basis modes in spatially
separated arms.

In order to characterize the output states of the proposed schemes, we
calculate their adjacency matrices, giving us the full
information about any quantum correlations contained in these
states. To do so, we make use of the fact that any unitary operation
$\hat U$ corresponding to a
symplectic transformation  $\hat U^\dag \,\mathbf{\hat x} \, \hat U =
\mathbf{S\,\hat x}$ changes the covariance matrix $\mathbf{C}$ of a
Gaussian state according to
\begin{equation}
\hat \rho \mapsto \hat U \,\hat \rho \,\hat U^\dag:\quad
\mathbf{C}^\rho \mapsto \mathbf{S\,C}^\rho \mathbf{S}^T .
\end{equation}
For a complete theoretical description of the state, however, we do
not only have to consider the co-rotating radially and azimuthally
polarized modes,
$\hat a_{R^+} = \hat a_R, ~ \hat a_{A^+} = \hat a_A$, but also their
counter-rotating complements, $\hat a_{A^-} = (\hat a_{V10} + \hat
a_{H01})/\sqrt{2} , ~ \hat a_{R^-} = (- \hat a_{H10} + \hat
a_{V01})/\sqrt{2}$ (Ref.~\cite{holleczek-classical-2011}.)
\begin{figure}
\begin{center}
\includegraphics[width=8cm]{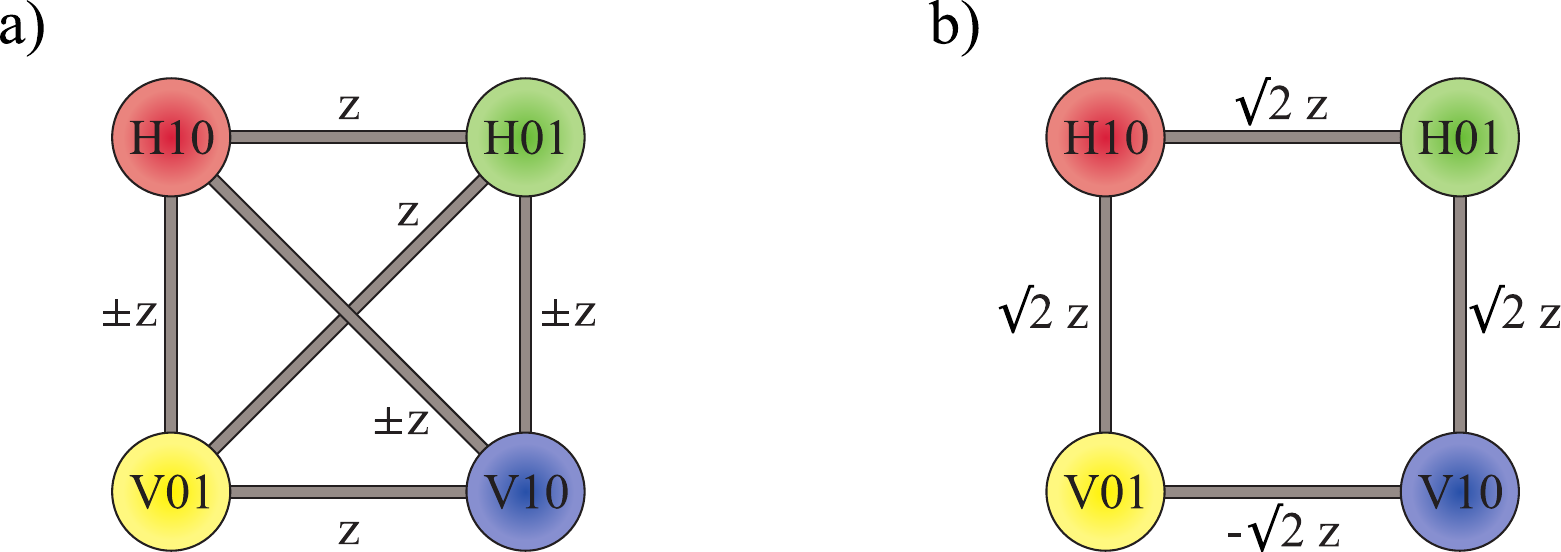}
\caption{\label{fig:Structure}a) The cluster states generated from scheme
  1, and b) from scheme 2. In scheme 1, a
  quadrature squeezed radially polarized mode is assumed as the input
  mode. The parameter $z$ is related to the squeezing of the
  cylindrically polarized mode at the input and, in these plots,
  quantifies the strength of the entanglement between the different,
  individually addressable  
  modes.}
\end{center}
\end{figure}

\begin{figure*}
\begin{center}
\includegraphics[width=16cm]{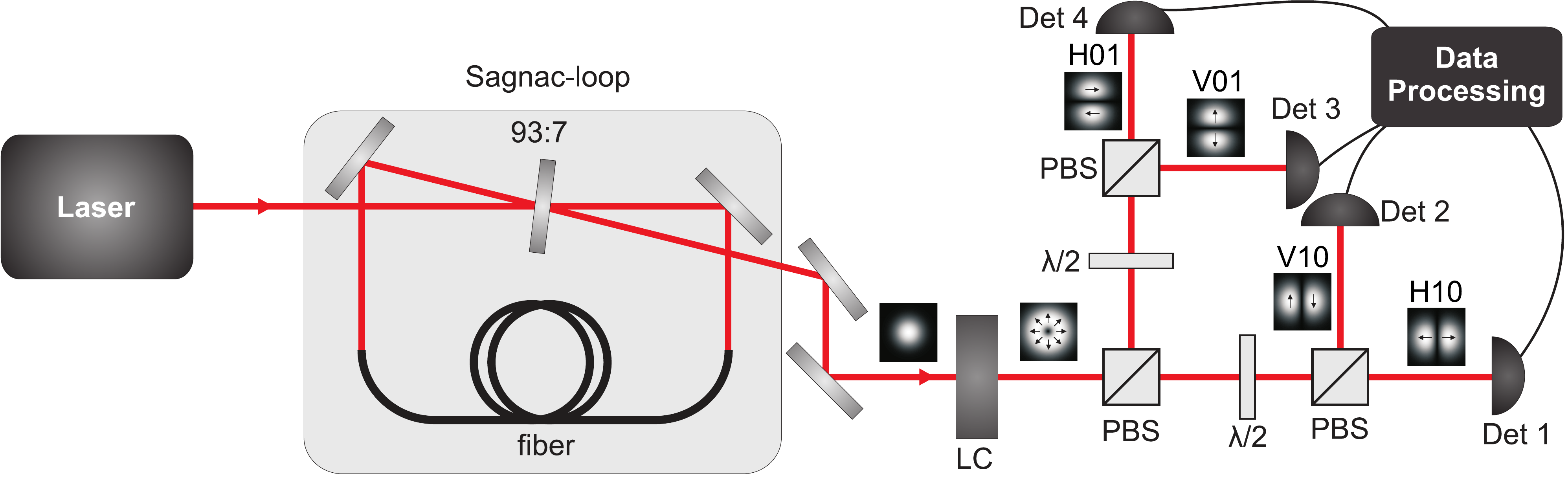}
\caption{\label{fig:Setup}The experimental setup to generate the
  cluster state from scheme 1 and measure the amplitude quantum
  correlations between the different modes. This scheme shows the case
  where a quadrature squeezed radially polarized beam enters the
  cluster-state generation circuit. By simply modifying the liquid
  crystal [LC], also an azimuthally polarized beam can be
  generated, while the passive circuit remains the same.
  [polarizing beam splitter (PBS), half-wave plate ($\lambda/2$)]}
\end{center}
\end{figure*}

The symplectic matrix for each  cluster generation scheme is composed
of the matrices of the individual squeezing and passive elements. From
the covariance matrix $\mathbf{C}^\mathrm{out }=\frac{1}{2}\,
\mathbf{S\,S}^T$, one can then extract the adjacency matrix of each
scheme according to (\ref{eq:DefUandV}).
As a final result for the adjacency matrices describing the cluster
graphs for schemes 1 and 2, we obtain
\begin{eqnarray}
  \label{eq:AdjacencyResult} \nonumber
  \mathbf{Z}_{1}^{A/R} &=& i\mathds{1}_4 + z \,\mathbf{\tilde
  V}_{1}^{A/R}, \\
  \mathbf{Z}_{2} &=& 2(z+i)\,\mathds{1}_4 + \sqrt{2}z\,\mathbf{\tilde V}_{2},
\end{eqnarray}
with
$$
\mathbf{\tilde V}_1^{A/R} =
\left(
  \begin{array}{cccc}
    1 & \pm 1 & 1 & \pm 1 \\
 \pm 1 & 1 & \pm 1 & 1 \\
 1 &  \pm 1 & 1 & \pm 1 \\
 \pm 1 & 1 & \pm 1 & 1
  \end{array}
\right)
 ,~
\mathbf{\tilde V}_{2}=
\left(
  \begin{array}{cccc}
    0 & 0 & 1 & 1\\
    0 & 0 & 1 & -1 \\
    1 & 1 & 0 & 0 \\
    1 & -1 & 0 & 0
\end{array}
\right)
 .
$$

Here, the first (last) two columns denote the beams with a
$\mathrm{TEM}_{10}$ ($\mathrm{TEM}_{01}$) spatial profile, while the
odd (even) number channels correspond to horizontal (vertical)
polarization.
The superscripts $A$ and $R$ represent the case for the radial or
azimuthal input mode in scheme 1, respectively. The parameter $z$
quantifies the amount of squeezing of the input modes. It is related
to the input squeezing in Eq.~(\ref{eq:Squeezer}) as $ z = (\langle
\hat q \hat p_\mathrm{in}
\rangle -i \langle \hat q^2_\mathrm{in} \rangle + \frac{i}{2}  )/4
\langle \hat q^2_\mathrm{in} \rangle$,  with $\langle \hat
q^2_\mathrm{in} \rangle$ and $\langle\hat q \hat p_\mathrm{in}
\rangle$ being the (co-)variances of the input mode. For a squeezed
input state, these two quantities are given by
\begin{eqnarray}
        \label{eq:InputSqueezing}
\nonumber
         \langle \hat q^2_\mathrm{in} \rangle  &=&
 \textstyle \frac{1}{2}\left[ \cosh
         (2r)  - \sinh(2r) \cos\theta\right] ,
\\
 \langle      \hat q \hat p_\mathrm{in} \rangle & =&
\textstyle
 -\frac{1}{2}\sinh(2r) \sin \theta .
      \end{eqnarray}
Here $r$ is the degree of squeezing and $\theta$ the squeezing
angle. The adjacency matrices clearly show that quantum correlations
between the different modes exist and that the output states form a
cluster. Especially intruiging is the form of the cluster's graph,
showing which modes are directly entangled to each other. The
topological structure of these states are displayed in
Fig.~\ref{fig:Structure}. The cluster state obtained from scheme 1 has
the unique feature that all modes are directly linked to each
other, corresponding to a so-called fully connected graph.
In the case of an azimuthally polarized input beam, it is even
\textit{fully symmetric} under permutation of its output modes. 
The cluster state from scheme 2 displays a box-like structure. Cluster states with this
graph topology have already proven their importance for quantum
teleportation protocols, both experimentally and
theoretically \cite{van-loock-building-2007,yukawa-experimental-2008}.

Our cluster-state generation schemes are distinct from other
schemes through the inclusion of the polarization degree of freedom. The
cylindrically polarized modes, which form the backbone of the setups, can be
manipulated by polarization optics and need only very few
interferences. Moreover, the extra polarization degree combined with the
spatial degree of freedom constitutes an interesting new information encoding
for quantum information processing.

As a proof of principle, we have verified that the cluster state setup
suggested in scheme 1 can actually be implemented. The experimental
realization is depicted in Fig.~\ref{fig:Setup}. A shot-noise limited
ORIGAMI laser (Onefive GmbH) emitting 220\,fs pulses centered at
1560\,nm acts as the light source.  The light is injected into an
asymmetric Sagnac interferometer \cite{schmitt-photon-number-1998} and generates an
amplitude squeezed linearly polarized Gaussian mode. The key
components of the Sagnac loop are a 93:7 beam splitter and a
$6.45\,\mbox{m}\pm0.02\,\mbox{m}$ long  polarization maintaining
single-mode fiber (3M FS-PM-7811) which acts as the nonlinear medium.
The Gaussian output mode displays an amplitude squeezing of
$-3.3\,\mathrm{dB}\pm0.1\,\mathrm{dB}$ at a sideband frequency of
$10.2$\,MHz. To realize scheme 1, one needs either a quadrature
squeezed azimuthally or radially polarized mode. The cylindrically
polarized mode is obtained by
mode transforming the Gaussian mode with the help of a liquid-crystal
polarization converter (ARCoptix) and wave plates
\cite{holleczek-classical-2011}. The 
losses at the device are $17\%\pm0.5\%$ and we have measured
$-1.9\,\mbox{dB}\pm0.1\,\mbox{dB}$ squeezing in both the radially and
azimuthally polarized
mode. In the next step, the cylindrically polarized mode goes through
a cascade of PBSs and half-wave plates, as indicated in
Fig. \ref{fig:Schemes}a. The four
outputs of the circuit are detected by single detectors
with sub-shot noise resolution at a sideband frequency of $10.2$\,MHz.

\begin{table}
\centering
\begin{tabular}{ | r |  r | r | }
  \hline
  {\bf Correlations} & {\bf Radial [dB]} & {\bf Azimuthal [dB]} \\
  \hline
  Var$\left\{I_\mathrm{H01}+I_\mathrm{H10}\right\}$ & $-0.8$ $(-0.8)$ & $-0.9$ $(-0.8)$\\
  \hline
  Var$\left\{I_\mathrm{H01}+I_\mathrm{V10}\right\}$ & $-0.8$ $(-0.8)$ & $-0.9$ $(-0.8)$\\
  \hline
  Var$\left\{I_\mathrm{V01}+I_\mathrm{H10}\right\}$ & $-0.8$ $(-0.8)$& $-0.8$ $(-0.8)$\\
  \hline
  Var$\left\{I_\mathrm{V01}+I_\mathrm{V10}\right\}$ & $-0.7$ $(-0.8)$& $-0.8$ $(-0.8)$\\
  \hline
  Var$\left\{I_\mathrm{H01}+I_\mathrm{V01}\right\}$ & $-0.8$ $(-0.8)$& $-0.9$ $(-0.8)$\\
  \hline
  Var$\left\{I_\mathrm{V10}+I_\mathrm{H10}\right\}$ & $-0.7$ $(-0.8)$& $-0.8$ $(-0.8)$\\
  \hline
  Var$\left\{I_\mathrm{H01}-I_\mathrm{H10}\right\}$ & $0.0$ $(0.0)$& $0.0$ $(0.0)$\\
  \hline
  Var$\left\{I_\mathrm{H01}-I_\mathrm{V10}\right\}$ & $0.0$ $(0.0)$& $0.0$ $(0.0)$\\
  \hline
  Var$\left\{I_\mathrm{V01}-I_\mathrm{H10}\right\}$ & $0.1$ $(0.0)$& $0.0$ $(0.0)$\\
  \hline
  Var$\left\{I_\mathrm{V01}-I_\mathrm{V10}\right\}$ & $0.1$ $(0.0)$& $0.0$ $(0.0)$\\
  \hline
  Var$\left\{I_\mathrm{H01}-I_\mathrm{V01}\right\}$ & $0.1$ $(0.0)$& $0.0$ $(0.0)$\\
  \hline
  Var$\left\{I_\mathrm{V10}-I_\mathrm{H10}\right\}$ & $0.0$ $(0.0)$& $0.0$ $(0.0)$\\
  \hline
\end{tabular}
\label{table:results}
\caption{Amplitude correlation measurements ($10.2$\,MHz sideband) between the different
  modes of the generated cluster states corresponding to scheme 1 for either
  the radially or the azimuthally polarized mode as the input mode. The
  expected theory value is displayed in brackets. The experimental error is 
  $\pm 0.1\,\mathrm{dB}$ in each case.}
\end{table}

The experimental scheme allows one to observe the amplitude quantum
correlations present between the different modes. Table
I lists all the different correlations and
anti-correlations between the different output modes and their
expected theoretical values. We find excellent agreement between
the theoretical and experimental results as expected due to the linearity and
simplicity of our setup. Stronger correlations can be achieved through higher
squeezing of the initial Gaussian mode or by squeezing the individual
orthogonal polarization modes as well as the two first-order Hermite Gaussian
modes. 

It should be noted that in order to fully characterize the 
system and directly verify the entanglement contained in the cluster,
measurements of the conjugate phases and all cross-correlations are also of
need. This is however not realizable with the present setup due to the
brightness of the resulting beams. However, as a first conceptual proof we may
infer from the predicted values for the amplitude correlations the proper
behavior of our passive setup.

In conclusion, we have demonstrated that quadrature squeezed
cylindrically polarized modes are ideal tools to generate CV cluster
states in a compact fashion.
The intrinsic entanglement contained within these modes allows
for a straightforward generation
of cluster states. As each node of the cluster state
is a unique mode, these can be easily addressed individually,
as needed for measurement-based quantum computations.
We have also shown that our approach is versatile and can be exploited to
obtain qualitatively different kinds of cluster states,
including a fully connected, fully symmetric graph of four optical modes.
The protocols proposed here clearly demonstrate that a
combination of different polarization states and higher-order spatial
modes, realizing a kind of hybrid entanglement
of different degrees of freedom, leads to compact and well addressable cluster
states. Our system is easily scalable by utilizing higher-order
$\mathrm{TEM}_{nm}$ modes, and the degree of correlation can be significantly
enhanced by squeezing each one of these modes by means of e.g.~multimode
parametric down-conversion. This approach may lead to further enhancement of
time-multiplexed schemes which have been used to create very
large cluster states \cite{yoshikawa-invited-2016}.

We would like to acknowledge financial support from the European Union
Integrating Project Q-ESSENCE. Peter van Loock wants to acknowledge
funding by an Emmy Noether Grant of the DFG.


\begin{thebibliography}{35}
\expandafter\ifx\csname natexlab\endcsname\relax\def\natexlab#1{#1}\fi
\expandafter\ifx\csname bibnamefont\endcsname\relax
  \def\bibnamefont#1{#1}\fi
\expandafter\ifx\csname bibfnamefont\endcsname\relax
  \def\bibfnamefont#1{#1}\fi
\expandafter\ifx\csname citenamefont\endcsname\relax
  \def\citenamefont#1{#1}\fi
\expandafter\ifx\csname url\endcsname\relax
  \def\url#1{\texttt{#1}}\fi
\expandafter\ifx\csname urlprefix\endcsname\relax\def\urlprefix{URL }\fi
\providecommand{\bibinfo}[2]{#2}
\providecommand{\eprint}[2][]{\url{#2}}

\bibitem[{\citenamefont{Bennett and DiVincenzo}(2000)}]{bennett-quantum-2000}
\bibinfo{author}{\bibfnamefont{C.~H.} \bibnamefont{Bennett}} \bibnamefont{and}
  \bibinfo{author}{\bibfnamefont{D.~P.} \bibnamefont{DiVincenzo}},
  \bibinfo{journal}{Nature} \textbf{\bibinfo{volume}{404}},
  \bibinfo{pages}{247} (\bibinfo{year}{2000}).

\bibitem[{\citenamefont{Raussendorf and
  Briegel}(2001)}]{raussendorf-one-way-2001}
\bibinfo{author}{\bibfnamefont{R.}~\bibnamefont{Raussendorf}} \bibnamefont{and}
  \bibinfo{author}{\bibfnamefont{H.~J.} \bibnamefont{Briegel}},
  \bibinfo{journal}{Phys.~Rev.~Lett.} \textbf{\bibinfo{volume}{86}},
  \bibinfo{pages}{5188} (\bibinfo{year}{2001}).

\bibitem[{\citenamefont{Raussendorf et~al.}(2003)\citenamefont{Raussendorf,
  Browne, and Briegel}}]{raussendorf-measurement-based-2003}
\bibinfo{author}{\bibfnamefont{R.}~\bibnamefont{Raussendorf}},
  \bibinfo{author}{\bibfnamefont{D.~E.} \bibnamefont{Browne}},
  \bibnamefont{and} \bibinfo{author}{\bibfnamefont{H.~J.}
  \bibnamefont{Briegel}}, \bibinfo{journal}{Phys.~Rev.~A}
  \textbf{\bibinfo{volume}{68}}, \bibinfo{pages}{022312}
  (\bibinfo{year}{2003}).

\bibitem[{\citenamefont{Nielsen}(2004)}]{nielsen-optical-2004}
\bibinfo{author}{\bibfnamefont{M.~A.} \bibnamefont{Nielsen}},
  \bibinfo{journal}{Phys.~Rev.~Lett.} \textbf{\bibinfo{volume}{93}},
  \bibinfo{pages}{040503} (\bibinfo{year}{2004}).

\bibitem[{\citenamefont{Walther et~al.}(2005)\citenamefont{Walther, Resch,
  Rudolph, Schenck, Weinfurter, Vedral, Aspelmeyer, and
  Zeilinger}}]{walther-experimental-2005}
\bibinfo{author}{\bibfnamefont{P.}~\bibnamefont{Walther}},
  \bibinfo{author}{\bibfnamefont{K.~J.} \bibnamefont{Resch}},
  \bibinfo{author}{\bibfnamefont{T.}~\bibnamefont{Rudolph}},
  \bibinfo{author}{\bibfnamefont{E.}~\bibnamefont{Schenck}},
  \bibinfo{author}{\bibfnamefont{H.}~\bibnamefont{Weinfurter}},
  \bibinfo{author}{\bibfnamefont{V.}~\bibnamefont{Vedral}},
  \bibinfo{author}{\bibfnamefont{M.}~\bibnamefont{Aspelmeyer}},
  \bibnamefont{and}
  \bibinfo{author}{\bibfnamefont{A.}~\bibnamefont{Zeilinger}},
  \bibinfo{journal}{Nature} \textbf{\bibinfo{volume}{434}},
  \bibinfo{pages}{169} (\bibinfo{year}{2005}).

\bibitem[{\citenamefont{Van~den Nest et~al.}(2006)\citenamefont{Van~den Nest,
  Miyake, Dür, and Briegel}}]{van-den-nest-universal-2006}
\bibinfo{author}{\bibfnamefont{M.}~\bibnamefont{Van~den Nest}},
  \bibinfo{author}{\bibfnamefont{A.}~\bibnamefont{Miyake}},
  \bibinfo{author}{\bibfnamefont{W.}~\bibnamefont{Dür}}, \bibnamefont{and}
  \bibinfo{author}{\bibfnamefont{H.}~\bibnamefont{Briegel}},
  \bibinfo{journal}{Phys.~Rev.~Lett.} \textbf{\bibinfo{volume}{97}}
  (\bibinfo{year}{2006}).

\bibitem[{\citenamefont{Menicucci et~al.}(2006)\citenamefont{Menicucci, van
  Loock, Gu, Weedbrook, Ralph, and Nielsen}}]{menicucci-universal-2006}
\bibinfo{author}{\bibfnamefont{N.~C.} \bibnamefont{Menicucci}},
  \bibinfo{author}{\bibfnamefont{P.}~\bibnamefont{van Loock}},
  \bibinfo{author}{\bibfnamefont{M.}~\bibnamefont{Gu}},
  \bibinfo{author}{\bibfnamefont{C.}~\bibnamefont{Weedbrook}},
  \bibinfo{author}{\bibfnamefont{T.~C.} \bibnamefont{Ralph}}, \bibnamefont{and}
  \bibinfo{author}{\bibfnamefont{M.~A.} \bibnamefont{Nielsen}},
  \bibinfo{journal}{Phys.~Rev.~Lett.} \textbf{\bibinfo{volume}{97}},
  \bibinfo{pages}{110501} (\bibinfo{year}{2006}).

\bibitem[{\citenamefont{Zhang and
  Braunstein}(2006)}]{zhang-continuous-variable-2006}
\bibinfo{author}{\bibfnamefont{J.}~\bibnamefont{Zhang}} \bibnamefont{and}
  \bibinfo{author}{\bibfnamefont{S.~L.} \bibnamefont{Braunstein}},
  \bibinfo{journal}{Phys.~Rev.~A} \textbf{\bibinfo{volume}{73}},
  \bibinfo{pages}{1} (\bibinfo{year}{2006}).

\bibitem[{\citenamefont{Su et~al.}(2007)\citenamefont{Su, Tan, Jia, Zhang, Xie,
  and Peng}}]{su-experimental-2007}
\bibinfo{author}{\bibfnamefont{X.}~\bibnamefont{Su}},
  \bibinfo{author}{\bibfnamefont{A.}~\bibnamefont{Tan}},
  \bibinfo{author}{\bibfnamefont{X.}~\bibnamefont{Jia}},
  \bibinfo{author}{\bibfnamefont{J.}~\bibnamefont{Zhang}},
  \bibinfo{author}{\bibfnamefont{C.}~\bibnamefont{Xie}}, \bibnamefont{and}
  \bibinfo{author}{\bibfnamefont{K.}~\bibnamefont{Peng}},
  \bibinfo{journal}{Phys.~Rev.~Lett.} \textbf{\bibinfo{volume}{98}},
  \bibinfo{pages}{070502} (\bibinfo{year}{2007}).

\bibitem[{\citenamefont{Menicucci et~al.}(2007)\citenamefont{Menicucci,
  Flammia, Zaidi, and Pfister}}]{menicucci-ultracompact-2007}
\bibinfo{author}{\bibfnamefont{N.~C.} \bibnamefont{Menicucci}},
  \bibinfo{author}{\bibfnamefont{S.~T.} \bibnamefont{Flammia}},
  \bibinfo{author}{\bibfnamefont{H.}~\bibnamefont{Zaidi}}, \bibnamefont{and}
  \bibinfo{author}{\bibfnamefont{O.}~\bibnamefont{Pfister}},
  \bibinfo{journal}{Phys.~Rev.~A} \textbf{\bibinfo{volume}{76}},
  \bibinfo{pages}{010302} (\bibinfo{year}{2007}).

\bibitem[{\citenamefont{van Loock et~al.}(2007)\citenamefont{van Loock,
  Weedbrook, and Gu}}]{van-loock-building-2007}
\bibinfo{author}{\bibfnamefont{P.}~\bibnamefont{van Loock}},
  \bibinfo{author}{\bibfnamefont{C.}~\bibnamefont{Weedbrook}},
  \bibnamefont{and} \bibinfo{author}{\bibfnamefont{M.}~\bibnamefont{Gu}},
  \bibinfo{journal}{Phys.~Rev.~A} \textbf{\bibinfo{volume}{76}},
  \bibinfo{pages}{1} (\bibinfo{year}{2007}).

\bibitem[{\citenamefont{Menicucci et~al.}(2008)\citenamefont{Menicucci,
  Flammia, and Pfister}}]{menicucci-one-way-2008}
\bibinfo{author}{\bibfnamefont{N.~C.} \bibnamefont{Menicucci}},
  \bibinfo{author}{\bibfnamefont{S.~T.} \bibnamefont{Flammia}},
  \bibnamefont{and} \bibinfo{author}{\bibfnamefont{O.}~\bibnamefont{Pfister}},
  \bibinfo{journal}{Phys.~Rev.~Lett.} \textbf{\bibinfo{volume}{101}},
  \bibinfo{pages}{1} (\bibinfo{year}{2008}).

\bibitem[{\citenamefont{Pysher et~al.}(2011)\citenamefont{Pysher, Miwa,
  Shahrokhshahi, Bloomer, and Pfister}}]{pysher-parallel-2011}
\bibinfo{author}{\bibfnamefont{M.}~\bibnamefont{Pysher}},
  \bibinfo{author}{\bibfnamefont{Y.}~\bibnamefont{Miwa}},
  \bibinfo{author}{\bibfnamefont{R.}~\bibnamefont{Shahrokhshahi}},
  \bibinfo{author}{\bibfnamefont{R.}~\bibnamefont{Bloomer}}, \bibnamefont{and}
  \bibinfo{author}{\bibfnamefont{O.}~\bibnamefont{Pfister}},
  \bibinfo{journal}{Phys.~Rev.~Lett.} \textbf{\bibinfo{volume}{107}},
  \bibinfo{pages}{030505} (\bibinfo{year}{2011}).

\bibitem[{\citenamefont{Menicucci et~al.}(2011)\citenamefont{Menicucci,
  Flammia, and van Loock}}]{menicucci-graphical-2011}
\bibinfo{author}{\bibfnamefont{N.~C.} \bibnamefont{Menicucci}},
  \bibinfo{author}{\bibfnamefont{S.~T.} \bibnamefont{Flammia}},
  \bibnamefont{and} \bibinfo{author}{\bibfnamefont{P.}~\bibnamefont{van
  Loock}}, \bibinfo{journal}{Phys.~Rev.~A} \textbf{\bibinfo{volume}{83}},
  \bibinfo{pages}{042335} (\bibinfo{year}{2011}).

\bibitem[{\citenamefont{Wang et~al.}(2014)\citenamefont{Wang, Chen, Menicucci,
  and Pfister}}]{wang-weaving-2014}
\bibinfo{author}{\bibfnamefont{P.}~\bibnamefont{Wang}},
  \bibinfo{author}{\bibfnamefont{M.}~\bibnamefont{Chen}},
  \bibinfo{author}{\bibfnamefont{N.~C.} \bibnamefont{Menicucci}},
  \bibnamefont{and} \bibinfo{author}{\bibfnamefont{O.}~\bibnamefont{Pfister}},
  \bibinfo{journal}{Phys.~Rev.~A} \textbf{\bibinfo{volume}{90}},
  \bibinfo{pages}{032325} (\bibinfo{year}{2014}).

\bibitem[{\citenamefont{Menicucci}(2014)}]{menicucci-fault-tolerant-2014}
\bibinfo{author}{\bibfnamefont{N.~C.} \bibnamefont{Menicucci}},
  \bibinfo{journal}{Phys.~Rev.~Lett.} \textbf{\bibinfo{volume}{112}},
  \bibinfo{pages}{120504} (\bibinfo{year}{2014}).

\bibitem[{\citenamefont{Roslund et~al.}(2013)\citenamefont{Roslund, de~Araújo,
  Jiang, Fabre, and Treps}}]{roslund-wavelength-multiplexed-2013}
\bibinfo{author}{\bibfnamefont{J.}~\bibnamefont{Roslund}},
  \bibinfo{author}{\bibfnamefont{R.~M.} \bibnamefont{de~Araújo}},
  \bibinfo{author}{\bibfnamefont{S.}~\bibnamefont{Jiang}},
  \bibinfo{author}{\bibfnamefont{C.}~\bibnamefont{Fabre}}, \bibnamefont{and}
  \bibinfo{author}{\bibfnamefont{N.}~\bibnamefont{Treps}},
  \bibinfo{journal}{Nature~Photon.} \textbf{\bibinfo{volume}{8}},
  \bibinfo{pages}{109} (\bibinfo{year}{2013}).

\bibitem[{\citenamefont{Alexander and
  Menicucci}(2016)}]{alexander-flexible-2016}
\bibinfo{author}{\bibfnamefont{R.~N.} \bibnamefont{Alexander}}
  \bibnamefont{and} \bibinfo{author}{\bibfnamefont{N.~C.}
  \bibnamefont{Menicucci}}, \bibinfo{journal}{Phys.~Rev.~A}
  \textbf{\bibinfo{volume}{93}}, \bibinfo{pages}{062326}
  (\bibinfo{year}{2016}).

\bibitem[{\citenamefont{Andersen et~al.}(2015)\citenamefont{Andersen,
  Neergaard-Nielsen, van Loock, and Furusawa}}]{andersen-hybrid-2015}
\bibinfo{author}{\bibfnamefont{U.~L.} \bibnamefont{Andersen}},
  \bibinfo{author}{\bibfnamefont{J.~S.} \bibnamefont{Neergaard-Nielsen}},
  \bibinfo{author}{\bibfnamefont{P.}~\bibnamefont{van Loock}},
  \bibnamefont{and} \bibinfo{author}{\bibfnamefont{A.}~\bibnamefont{Furusawa}},
  \bibinfo{journal}{Nature~Phys.} \textbf{\bibinfo{volume}{11}},
  \bibinfo{pages}{713} (\bibinfo{year}{2015}).

\bibitem[{\citenamefont{Yukawa et~al.}(2008)\citenamefont{Yukawa, Ukai, van
  Loock, and Furusawa}}]{yukawa-experimental-2008}
\bibinfo{author}{\bibfnamefont{M.}~\bibnamefont{Yukawa}},
  \bibinfo{author}{\bibfnamefont{R.}~\bibnamefont{Ukai}},
  \bibinfo{author}{\bibfnamefont{P.}~\bibnamefont{van Loock}},
  \bibnamefont{and} \bibinfo{author}{\bibfnamefont{A.}~\bibnamefont{Furusawa}},
  \bibinfo{journal}{Phys.~Rev.~A} \textbf{\bibinfo{volume}{78}},
  \bibinfo{pages}{012301} (\bibinfo{year}{2008}).

\bibitem[{\citenamefont{Armstrong et~al.}(2012)\citenamefont{Armstrong,
  Morizur, Janousek, Hage, Treps, Lam, and
  Bachor}}]{armstrong-programmable-2012}
\bibinfo{author}{\bibfnamefont{S.}~\bibnamefont{Armstrong}},
  \bibinfo{author}{\bibfnamefont{J.-F.} \bibnamefont{Morizur}},
  \bibinfo{author}{\bibfnamefont{J.}~\bibnamefont{Janousek}},
  \bibinfo{author}{\bibfnamefont{B.}~\bibnamefont{Hage}},
  \bibinfo{author}{\bibfnamefont{N.}~\bibnamefont{Treps}},
  \bibinfo{author}{\bibfnamefont{P.~K.} \bibnamefont{Lam}}, \bibnamefont{and}
  \bibinfo{author}{\bibfnamefont{H.-A.} \bibnamefont{Bachor}},
  \bibinfo{journal}{Nature~Commun.} \textbf{\bibinfo{volume}{3}},
  \bibinfo{pages}{1026} (\bibinfo{year}{2012}).

\bibitem[{\citenamefont{Pinel et~al.}(2012)\citenamefont{Pinel, Jian,
  de~Araújo, Feng, Chalopin, Fabre, and Treps}}]{pinel-generation-2012}
\bibinfo{author}{\bibfnamefont{O.}~\bibnamefont{Pinel}},
  \bibinfo{author}{\bibfnamefont{P.}~\bibnamefont{Jian}},
  \bibinfo{author}{\bibfnamefont{R.}~\bibnamefont{de~Araújo}},
  \bibinfo{author}{\bibfnamefont{J.}~\bibnamefont{Feng}},
  \bibinfo{author}{\bibfnamefont{B.}~\bibnamefont{Chalopin}},
  \bibinfo{author}{\bibfnamefont{C.}~\bibnamefont{Fabre}}, \bibnamefont{and}
  \bibinfo{author}{\bibfnamefont{N.}~\bibnamefont{Treps}},
  \bibinfo{journal}{Phys.~Rev.~Lett.} \textbf{\bibinfo{volume}{108}},
  \bibinfo{pages}{22} (\bibinfo{year}{2012}).

\bibitem[{\citenamefont{Menicucci et~al.}(2010)\citenamefont{Menicucci, Ma, and
  Ralph}}]{menicucci-arbitrarily-2010}
\bibinfo{author}{\bibfnamefont{N.~C.} \bibnamefont{Menicucci}},
  \bibinfo{author}{\bibfnamefont{X.}~\bibnamefont{Ma}}, \bibnamefont{and}
  \bibinfo{author}{\bibfnamefont{T.~C.} \bibnamefont{Ralph}},
  \bibinfo{journal}{Phys.~Rev.~Lett.} \textbf{\bibinfo{volume}{104}},
  \bibinfo{pages}{250503} (\bibinfo{year}{2010}).

\bibitem[{\citenamefont{Menicucci}(2011)}]{menicucci-temporal-mode-2011}
\bibinfo{author}{\bibfnamefont{N.~C.} \bibnamefont{Menicucci}},
  \bibinfo{journal}{Phys.~Rev.~A} \textbf{\bibinfo{volume}{83}},
  \bibinfo{pages}{062314} (\bibinfo{year}{2011}).

\bibitem[{\citenamefont{Chen et~al.}(2014)\citenamefont{Chen, Menicucci, and
  Pfister}}]{chen-experimental-2014}
\bibinfo{author}{\bibfnamefont{M.}~\bibnamefont{Chen}},
  \bibinfo{author}{\bibfnamefont{N.~C.} \bibnamefont{Menicucci}},
  \bibnamefont{and} \bibinfo{author}{\bibfnamefont{O.}~\bibnamefont{Pfister}},
  \bibinfo{journal}{Phys.~Rev.~Lett.} \textbf{\bibinfo{volume}{112}},
  \bibinfo{pages}{120505} (\bibinfo{year}{2014}).

\bibitem[{\citenamefont{Yokoyama et~al.}(2013)\citenamefont{Yokoyama, Ukai,
  Armstrong, Sornphiphatphong, Kaji, Suzuki, Yoshikawa, Yonezawa, Menicucci,
  and Furusawa}}]{yokoyama-ultra-large-scale-2013}
\bibinfo{author}{\bibfnamefont{S.}~\bibnamefont{Yokoyama}},
  \bibinfo{author}{\bibfnamefont{R.}~\bibnamefont{Ukai}},
  \bibinfo{author}{\bibfnamefont{S.~C.} \bibnamefont{Armstrong}},
  \bibinfo{author}{\bibfnamefont{C.}~\bibnamefont{Sornphiphatphong}},
  \bibinfo{author}{\bibfnamefont{T.}~\bibnamefont{Kaji}},
  \bibinfo{author}{\bibfnamefont{S.}~\bibnamefont{Suzuki}},
  \bibinfo{author}{\bibfnamefont{J.-I.} \bibnamefont{Yoshikawa}},
  \bibinfo{author}{\bibfnamefont{H.}~\bibnamefont{Yonezawa}},
  \bibinfo{author}{\bibfnamefont{N.~C.} \bibnamefont{Menicucci}},
  \bibnamefont{and} \bibinfo{author}{\bibfnamefont{A.}~\bibnamefont{Furusawa}},
  \bibinfo{journal}{Nature~Photon.} \textbf{\bibinfo{volume}{7}},
  \bibinfo{pages}{982} (\bibinfo{year}{2013}).

\bibitem[{\citenamefont{Yoshikawa et~al.}(2016)\citenamefont{Yoshikawa,
  Yokoyama, Kaji, Sornphiphatphong, Shiozawa, Makino, and
  Furusawa}}]{yoshikawa-invited-2016}
\bibinfo{author}{\bibfnamefont{J.-I.} \bibnamefont{Yoshikawa}},
  \bibinfo{author}{\bibfnamefont{S.}~\bibnamefont{Yokoyama}},
  \bibinfo{author}{\bibfnamefont{T.}~\bibnamefont{Kaji}},
  \bibinfo{author}{\bibfnamefont{C.}~\bibnamefont{Sornphiphatphong}},
  \bibinfo{author}{\bibfnamefont{Y.}~\bibnamefont{Shiozawa}},
  \bibinfo{author}{\bibfnamefont{K.}~\bibnamefont{Makino}}, \bibnamefont{and}
  \bibinfo{author}{\bibfnamefont{A.}~\bibnamefont{Furusawa}},
  \bibinfo{journal}{Appl.~Phys.~Lett.~Photon.} \textbf{\bibinfo{volume}{1}},
  \bibinfo{pages}{060801} (\bibinfo{year}{2016}).

\bibitem[{\citenamefont{Gabriel et~al.}(2011)\citenamefont{Gabriel, Aiello,
  Zhong, Euser, Joly, Banzer, Förtsch, Elser, Andersen, Marquardt
  et~al.}}]{gabriel-entangling-2011}
\bibinfo{author}{\bibfnamefont{C.}~\bibnamefont{Gabriel}},
  \bibinfo{author}{\bibfnamefont{A.}~\bibnamefont{Aiello}},
  \bibinfo{author}{\bibfnamefont{W.}~\bibnamefont{Zhong}},
  \bibinfo{author}{\bibfnamefont{T.~G.} \bibnamefont{Euser}},
  \bibinfo{author}{\bibfnamefont{N.~Y.} \bibnamefont{Joly}},
  \bibinfo{author}{\bibfnamefont{P.}~\bibnamefont{Banzer}},
  \bibinfo{author}{\bibfnamefont{M.}~\bibnamefont{Förtsch}},
  \bibinfo{author}{\bibfnamefont{D.}~\bibnamefont{Elser}},
  \bibinfo{author}{\bibfnamefont{U.~L.} \bibnamefont{Andersen}},
  \bibinfo{author}{\bibfnamefont{C.}~\bibnamefont{Marquardt}},
  \bibnamefont{et~al.}, \bibinfo{journal}{Phys.~Rev.~Lett.}
  \textbf{\bibinfo{volume}{106}}, \bibinfo{pages}{1} (\bibinfo{year}{2011}).

\bibitem[{\citenamefont{Aiello et~al.}(2015)\citenamefont{Aiello, Töppel,
  Marquardt, Giacobino, and Leuchs}}]{aiello-quantumlike-2015}
\bibinfo{author}{\bibfnamefont{A.}~\bibnamefont{Aiello}},
  \bibinfo{author}{\bibfnamefont{F.}~\bibnamefont{Töppel}},
  \bibinfo{author}{\bibfnamefont{C.}~\bibnamefont{Marquardt}},
  \bibinfo{author}{\bibfnamefont{E.}~\bibnamefont{Giacobino}},
  \bibnamefont{and} \bibinfo{author}{\bibfnamefont{G.}~\bibnamefont{Leuchs}},
  \bibinfo{journal}{New~J.~Phys.} \textbf{\bibinfo{volume}{17}},
  \bibinfo{pages}{043024} (\bibinfo{year}{2015}).

\bibitem[{\citenamefont{dos Santos et~al.}(2009)\citenamefont{dos Santos,
  Dechoum, and Khoury}}]{dos-santos-continuous-variable-2009}
\bibinfo{author}{\bibfnamefont{B.~C.} \bibnamefont{dos Santos}},
  \bibinfo{author}{\bibfnamefont{K.}~\bibnamefont{Dechoum}}, \bibnamefont{and}
  \bibinfo{author}{\bibfnamefont{A.~Z.} \bibnamefont{Khoury}},
  \bibinfo{journal}{Phys.~Rev.~Lett.} \textbf{\bibinfo{volume}{103}}
  (\bibinfo{year}{2009}).

\bibitem[{\citenamefont{Lassen et~al.}(2009)\citenamefont{Lassen, Leuchs, and
  Andersen}}]{lassen-continuous-2009}
\bibinfo{author}{\bibfnamefont{M.}~\bibnamefont{Lassen}},
  \bibinfo{author}{\bibfnamefont{G.}~\bibnamefont{Leuchs}}, \bibnamefont{and}
  \bibinfo{author}{\bibfnamefont{U.~L.} \bibnamefont{Andersen}},
  \bibinfo{journal}{Phys.~Rev.~Lett.} \textbf{\bibinfo{volume}{102}},
  \bibinfo{pages}{163602} (\bibinfo{year}{2009}).

\bibitem[{\citenamefont{Euser et~al.}(2011)\citenamefont{Euser, Schmidt, Joly,
  and Gabriel}}]{euser-birefringence-2011}
\bibinfo{author}{\bibfnamefont{T.~G.} \bibnamefont{Euser}},
  \bibinfo{author}{\bibfnamefont{M.}~\bibnamefont{Schmidt}},
  \bibinfo{author}{\bibfnamefont{N.~Y.} \bibnamefont{Joly}}, \bibnamefont{and}
  \bibinfo{author}{\bibfnamefont{C.}~\bibnamefont{Gabriel}},
  \bibinfo{journal}{J.~Opt.~Soc.~A.~B} \textbf{\bibinfo{volume}{28}},
  \bibinfo{pages}{193} (\bibinfo{year}{2011}).

\bibitem[{\citenamefont{Briegel and
  Raussendorf}(2001)}]{briegel-persistent-2001}
\bibinfo{author}{\bibfnamefont{H.~J.} \bibnamefont{Briegel}} \bibnamefont{and}
  \bibinfo{author}{\bibfnamefont{R.}~\bibnamefont{Raussendorf}},
  \bibinfo{journal}{Phys.~Rev.~Lett.} \textbf{\bibinfo{volume}{86}},
  \bibinfo{pages}{910} (\bibinfo{year}{2001}).

\bibitem[{\citenamefont{Holleczek et~al.}(2011)\citenamefont{Holleczek, Aiello,
  Gabriel, Marquardt, and Leuchs}}]{holleczek-classical-2011}
\bibinfo{author}{\bibfnamefont{A.}~\bibnamefont{Holleczek}},
  \bibinfo{author}{\bibfnamefont{A.}~\bibnamefont{Aiello}},
  \bibinfo{author}{\bibfnamefont{C.}~\bibnamefont{Gabriel}},
  \bibinfo{author}{\bibfnamefont{C.}~\bibnamefont{Marquardt}},
  \bibnamefont{and} \bibinfo{author}{\bibfnamefont{G.}~\bibnamefont{Leuchs}},
  \bibinfo{journal}{Opt.~Express} \textbf{\bibinfo{volume}{19}},
  \bibinfo{pages}{9714} (\bibinfo{year}{2011}).

\bibitem[{\citenamefont{Schmitt et~al.}(1998)\citenamefont{Schmitt, Ficker,
  Wolff, König, Sizmann, and Leuchs}}]{schmitt-photon-number-1998}
\bibinfo{author}{\bibfnamefont{S.}~\bibnamefont{Schmitt}},
  \bibinfo{author}{\bibfnamefont{J.}~\bibnamefont{Ficker}},
  \bibinfo{author}{\bibfnamefont{M.}~\bibnamefont{Wolff}},
  \bibinfo{author}{\bibfnamefont{F.}~\bibnamefont{König}},
  \bibinfo{author}{\bibfnamefont{A.}~\bibnamefont{Sizmann}}, \bibnamefont{and}
  \bibinfo{author}{\bibfnamefont{G.}~\bibnamefont{Leuchs}},
  \bibinfo{journal}{Phys.~Rev.~Lett.} \textbf{\bibinfo{volume}{81}},
  \bibinfo{pages}{2446} (\bibinfo{year}{1998}).

\end{thebibliography}
\end{document}